\documentclass{jfm}
\usepackage{fancyhdr}
\usepackage{graphicx}
\usepackage{epstopdf, epsfig}
\usepackage{dcolumn}
\usepackage{bm}
\usepackage{pgfplots}
\usetikzlibrary{matrix}
\usepackage{tikz}
\usepackage{caption}
\usepackage{subcaption}
\pgfplotsset{compat = newest}
\usepackage{subcaption}
\usepackage{mathptmx}
\usepackage{helvet}
\usepackage{etoolbox}
\usepackage{amsmath}
\usepackage{xcolor}
\usepackage[colorlinks=true]{hyperref}
\usepackage{appendix}

\newcommand{\ba}[1]{{\bm{a}}^{(#1)}}

\newcommand{\bx}{\bm{x}}
\newcommand{\bc}{\bm{c}}
\newcommand{\bxi}{\bm{\xi}}
\newcommand{\bg}{\bm{g}}

\newcommand{\bu}{\bm{u}}

\newcommand{\bsigma}{\bm{\sigma}}
\newcommand{\bdelta}{\bm{\delta}}
\newcommand{\f}[1]{f^{(#1)}}

\newcommand{\bH}[1]{{\cal H}^{(#1)}}

\shorttitle{LBM Forcing scheme for thermal compressible flows}
\shortauthor{Zuoxu Li and Xiaowen Shan}

\title{Body-force modelling in thermal compressible flows with lattice Boltzmann
	method}

\author{Zuoxu Li \and Xiaowen Shan\corresp{\email{shanxw@sustech.edu.cn}}}

\affiliation{Department of Mechanics and Aerospace Engineering, Southern
	University of Science and Technology, Shenchen 518055, China}

\begin{document}

\maketitle

\begin{abstract}
Body force modelling in lattice Boltzmann method (LBM) has been extensively
studied in the incompressible limit but rarely discussed for thermal
compressible flows. Here we present a systematic approach of incorporating body
force in LBM which is valid for thermal compressible and non-equilibrium flows.
In particular, a LBM forcing scheme accurate for the energy equation with
second-order time accuracy is given.  New and essential in this scheme is the
third-moment contribution of the force term. It is shown via Chapman-Enskog
analysis that the absence of this contribution causes an erroneous heat flux
quadratic in Mach number and linear in temperature variation. The theoretical
findings are verified and the necessity of the third-moment contribution
demonstrated by numerical simulations.
\end{abstract}

\begin{keywords}
Lattice Boltzmann method
\end{keywords}

\section{Introduction}

Flows involving body forces are widely present in nature and engineering
practice.  Examples include gravity induced natural convection such as in
Rayleigh-Bernard flows, flows in rotating reference system influenced by
centrifugal and Coriolis forces, magneto-fluid influenced by electromagnetic
force, and many others.  In the fast-growing lattice Boltzmann method (LBM),
correctly incorporating the body force has an added importance as the body force
is also used to model inter-particle interactions giving rise to the rich
phenomena of multiphase flows~\citep{Shan1993,He1998a}.

Although the treatment of body force is rather straightforward in both
hydrodynamic equations and Boltzmann equation, it remains a non-trivial and
sometimes even controversial topic for LBM after a substantial amount of effort
and literally a dozen proposed schemes~\citep{Mohamad2010a,Bawazeer2021}. The
difficulty can be arguably attributed to the fact that LBM was developed from
beginning as a discrete kinetic model that was tailored \textit{a posteriori} to
exhibit Navier-Stokes-Fourier thermohydrodynamics at the macroscopic level. It
lacked a first-principle theory dictating the evolution of the discrete
distribution in an external force field. Furthermore, from the perspective of
kinetic theory, the configuration-velocity phase space was discretized together,
making error analysis complicated.

An early idea was suggested by~\cite{Shan1993} to shift the equilibrium velocity
in the collision term to account for the change of momentum which is the
leading-order effect of body force. This approach only imposes a condition on
the zeroth and first moments of the discrete distribution function. Although
reasonably successful in simple flows, subtle issues arose in applications where
detailed modeling of the body force effect is required. An example is the
multi-phase fluid where the velocity-shift scheme results in unphysical
dependence of the equilibrium densities on the relaxation time~\citep{Yu2009}.
\cite{Sbragaglia2009c} eliminated the inaccuracy in energy equation  generated
by the shifted velocity by also shifting the temperature. Using the method of
undetermined coefficients and matching the macroscopic equation out of
Chapman-Enskog calculation with the Navier-Stokes equation, \cite{Guo2002} gave
a scheme to eliminate the discrete lattice effect caused by spatial-temporal
discretization, which also removed the unphysical dependency on the
relaxation-time when applied to the multiphase model. This line of approaches,
see also~\cite{Buick2000a,Ladd2001}, calls upon \textit{a posteriori} matching
with the hydrodynamic equations and becomes unwieldly when applied to more
complicated collision models and higher-order hydrodynamic approximations.

Another strategy is to start from the body force term of the Boltzmann equation,
$\bg\cdot\nabla_{\xi} f$. \cite{He1998} approximated $f$ by the
Maxwell-Boltzmann equilibrium, $\f{0}$, and integrated using the trapezoidal
rule to advance one time step. An auxiliary variable was introduced to eliminate
the implicity. \cite{Martys1998} pointed out, by examining the Hermite expansion
of the body force, that this approximation is exact up to the second moments. By
realizing that $\nabla_\xi\f{0} = -\nabla_u\f{0}$, \cite{Kupershtokh2004}
introduced the exact difference method (EDM) to model the effect of the body
force as $\f{0}(\rho, \bu+\bg\Delta t) - \f{0}(\rho, \bu)$ which correctly
advances a local equilibrium to another one with a shifted velocity. However,
effect of the non-equilibrium part of the distribution is ignored all together.
In both schemes the spatial-temporal discretization was carefully handled to
achieve second-order accuracy.

Despite that the differences, similarities, and accuracies of the existing
forcing schemes have been theoretically analysed and numerically examined by a
number of
authors~\citep{Kupershtokh2009,Mohamad2010a,Huang2011d,Silva2012,Li2012b}, it
remains inconclusive as to which scheme is the most accurate because the results
often depend on flow conditions such as compressibility and steadiness. Possible
origins of the discrepancies include insufficient expansions of the distribution
function, the collision term, or the body force term, all resulting in different
error terms in the recovered macroscopic equations.  More importantly, as the
LBM being extended to thermal compressible flows, the existing works mainly
focus on the recovery of the mass and momentum equations, whereas, except for a
few works~\citep{Sbragaglia2009c}, the body force effect on the energy equation
is rarely discussed.

More recently the LBM was re-formulated using kinetic theory in continuum in two
separate steps~\citep{Shan1998,Shan2006b}.  First, the BGK model equation is
projected into a finite-dimensional Hilbert space spanned by the Hermite
polynomials. By choosing a set of discrete velocity coordinates that forms a
Gauss-Hermite quadrature rule and evaluating the BGK equation at these
velocities, one obtains the dynamic equations for a set of discrete
distributions in the configuration space which preserve the moments of the
continuum distribution function and therefore the hydrodynamics of the BGK
equation. Second, the equations for the discrete distribution are further
discretized in space and time to yield the lattice Boltzmann equation.  With the
velocity-space discretization of the body force term simply amounted to
expanding it in Hermite polynomials, \cite{Li2022} showed that by using
second-order time integration on the second-order Hermite expansion of the body
force term, one can recover \textit{a priori} the force scheme of
\cite{Guo2002}.  Moreover, the methodology is generic so that it can be applied
to higher-order moment expansions which give rise to the thermal and
non-equilibrium dynamics.

In this paper we present a systematic discretization scheme for the body force
term which is second-order in space and time, and valid for moments of arbitrary
orders.  In particular, force terms pertinent to the energy equation are
explicitly given, and errors in heat flux caused by insufficient expansion of
the force term is also obtained \textit{via} Chapman-Enskog analysis.  Numerical
verifications were carried out to show that the third-moment contribution in the
force term has a non-negligible effect in flows with strong temperature
variation. The remainder of the paper is organized as follows. In
Section~\ref{sec:theory}, we present the methodology to systematically obtain
the force term in LBM. The third-order expansion of the form term is obtained
and its necessity in the energy equation is demonstrated in Chapman-Enskog
analysis. In Section~\ref{sec:numerical}, we present numerical verifications and
analyses. Some further discussions are given in Section~\ref{sec:discussion},

\section{Moment expansion of the force term}

\label{sec:theory}

We first briefly recap the kinetic theoretic formulation of the LBM. Start with
the Boltzmann equation with the BGK collision operator
\begin{align}
    \label{eq:BGK}
     \frac{\p f}{\p t} +  \bm{\xi}\cdot\nabla f + \bm{g}\cdot\nabla_{\xi} f = \Omega(f)
    = -\frac{1}{\tau}\left[f-\f{0}\right],
\end{align}
where $\bx$ and $\bxi$ are respectively the coordinates in physical and velocity
spaces, $t$ the time, $f(\bm{x},\bm{\xi},t)$ the single-particle distribution
function, $\bm{g}$ the acceleration of the body force, $\nabla_ {\xi}$ the
gradient in velocity space, $\Omega$ the collision term, $\tau$ the relaxation
time, and $\f{0}$ the Maxwell-Boltzmann equilibrium distribution function:
\begin{equation}
	\f{0} = \frac{\rho}{(2\pi)^{D/2}}\exp\left[-\frac{c^2}{2\theta}\right],
\end{equation}
where $D$ is the space dimensionality, $\bm{c}=\bm{\xi}-\bm{u}$ and $c^2 =
\bm{c}\cdot\bm{c}$. The fluid density, $\rho$, velocity, $\bu$, and temperature,
$\theta$, are velocity moments of the distribution function
\begin{align}
	\label{eq:rut}
    \left\{\rho, \rho\bu, D\rho\theta\right\} =
    \int f \left\{1, \bm{\xi}, c^2\right\}d\bm{\xi}.
\end{align} 
The LBM can be formulated as a velocity-space discretization by first projecting
Eq.~\eqref{eq:BGK} into a finite-dimensional functional space spanned by Hermite
polynomials and then evaluating at discrete velocities that form a Gauss-Hermite
quadrature rule in the velocity space~\citep{Grad1949a,Shan1998,Shan2006b}. The
latter requirement ensures that the leading velocity moments are exactly
preserved by the discrete distribution function values. A key step in this
formulation is to expand all terms in Eq.~\eqref{eq:BGK} into finite Hermite
series. Assuming $f$ has the following expansion in the laboratory reference
frame
\begin{equation}
	\label{eq:fN}
	f_N(\bx, \bxi, t) = \omega(\bxi)\sum_{n=0}^N\frac{1}{n!}
	\ba{n}(\bx, t):\bH{n}(\bxi),
\end{equation}
where, $N$ is the expansion order, `:' stands for full contraction between two
tensors, and $\ba{n}(\bx,t)$ the expansion coefficients given by
\begin{equation}
	\label{eq:an}
	\ba{n}(\bx,t) = \int f_N(\bm{x},\bm{\xi},t)\bH{n}(\bxi)d\bxi.
\end{equation}
The corresponding expansion of the body-force term was given
by~\cite{Martys1998} as
\begin{equation}
	\label{eq:F}
	F(\bxi)\equiv -\bg\cdot\nabla_{\xi}f_N = \omega(\bxi)\sum_{n=1}^{N+1}
	\frac{1}{n!}n\bg\ba{n-1}:\bH{n}(\bxi),
\end{equation}
where $\bg\ba{n-1}$ is a rank-$n$ symmetric tensor denoting the
\textit{symmetric product} between $\bg$ and $\ba{n-1}$. Noticing that $\ba{0} =
\rho$ and $\ba{1} = \rho\bu$, the first two terms in Eq.~\eqref{eq:F} can be
evaluated as
\begin{equation}
	F(\bxi) \cong \omega\rho\left[\bg\cdot\bxi
	+ (\bg\cdot\bxi)(\bu\cdot\bxi) - \bg\cdot\bu\right].
\end{equation}
We note that most existing force schemes are based on the expression
above~\citep{Li2022}. To extend the body force to higher moments, the higher
terms can be calculated using Eq.~\eqref{eq:an}, \textit{e.g.}
\begin{equation}
	\label{eq:f2}
	\ba{2} = \int f_N\left(\bxi^2 - \bdelta\right )d\bxi
	= \rho\left(\bu^2-\bdelta\right) + \int f_N\bc\bc d\bxi.
\end{equation}
The last term is the \textit{momentum flux density} tensor which can be
decomposed into the \textit{normal pressure}, $\rho\bdelta$, and the traceless
\textit{deviatoric stress} tensor, $\bsigma$, as
\begin{equation}
	\int f_N\bc\bc d\bxi = p\bdelta -\bsigma.
\end{equation}
Comparing with the Hermite expansion of $\f{0}$, we recognize that
\begin{equation}
	\ba{2} = \ba{2}_0 - \bsigma,
\end{equation}
where $\ba{n}_0$ denotes the Hermite coefficients of $\f{0}$. Using $\bH{3} =
\bxi^3 - 3\bxi\bdelta$, the third-order term can be obtained as
\begin{eqnarray}
   		\label{eq:f3}
		\bg\ba{2}:\bH{3} &=& \rho\left\{(\bg\cdot\bxi)
		\left[(\bu\cdot\bxi)^2 -u^2 + (\theta-1)(\xi^2-D-2)\right]
		-2(\bg\cdot\bu)(\bu\cdot\bxi)\right\} \nonumber \\
		&&- (\bg\cdot\bxi)\bsigma:\bxi\bxi + 2\bsigma:\bg\bxi. 
\end{eqnarray}

Two remarks can be made here. First, the first term on the \textit{r.h.s.}\ is
the contribution of $\ba{2}_0$ and is proportional to $u^2$ or $\theta-1$. The
terms proportional to $\bsigma$ are the contributions of the non-equilibrium
part of the distribution.  Secondly, the body force generates momentum and total
kinetic energy in the form of bulk fluid motion, but does not generate any heat
as can be seen from its heat production rate
\begin{equation}
	\int\left(\bg\cdot\nabla_\xi f\right)c^2d\bxi = \bg\cdot\int c^2\nabla_\xi fd\bxi.
\end{equation}
Integrating by part and noticing that as $\xi\rightarrow\infty$, $f$ vanishes
faster than any power of $\xi$, we have
\begin{equation}
\label{eq:F-heat}
    \int c^2\nabla_\xi fd\bxi = -\int f\nabla_{\xi}c^2d\bxi 
     = -2\int f\left(\bxi-\bu\right)d\bxi = 0.
\end{equation}
Note that for the truncated $f_N$, the above equalities hold if and only if
$N\geq 2$.

Once restricted in the finite-dimensional Hermite space, Eq.~\eqref{eq:BGK} can
be further discretized to yield the lattice Boltzmann
equations~\citep{Shan2006b}. Let $w_i$ and $\bxi_i$, $i\in\{1, \cdots, d\}$, be
the weights and abscissas of a degree-$Q$ Gauss-Hermite quadrature rule,
\textit{i.e.} for any polynomial, $p(\bxi)$, of a degree not exceeding $Q$, we
have
\begin{equation}
	\int\omega(\bxi)p(\bxi)d\bxi = \sum_{i=1}^dw_ip(\bxi_i).
\end{equation}
For the $f_N$ in Eq.~\eqref{eq:fN}, $f_N/\omega$ is a degree-$N$ polynomial.
Provided that $Q\geq 2N$, Eq.~\eqref{eq:an} reduces to a summation
\begin{equation}
	\label{eq:an1}
	\ba{n} = \int\omega\left[\frac{f_N}{\omega}\bH{n}\right]d\bxi
	=\sum_{i=1}^df_i\bH{n}(\bxi_i),\quad\forall n\leq N,
\end{equation}
where $f_i$ is the \textit{discrete distribution} defined as
\begin{equation}
	f_i(\bx, t)\equiv \frac{w_if_N(\bx, \bxi_i, t)}{\omega(\bxi_i)},\quad
	i \in\left\{1, \cdots, d\right\}.
\end{equation}
In particular, as special cases of Eq.~\eqref{eq:an1}, Eqs.~\eqref{eq:rut}
become
\begin{align}
	\label{eq:rut1}
	\rho = \sum_if_i,\quad
	\rho\bu = \sum_if_i\bxi_i,\quad
	\rho\left(u^2 + D\theta\right) = \sum_if_i\xi_i^2.
\end{align}
The governing equation of $f_i$ can be obtained by evaluating Eq.~\eqref{eq:BGK} at
$\bxi_i$:
\begin{align}
	\label{eq:dbe}
	\frac{\p f_i}{\p t} + \bxi_i\cdot\nabla f_i = \Omega_i(f) + F_i,\quad
	i \in\left\{1, \cdots, d\right\}.
\end{align}
where $F_i \equiv w_iF(\bxi_i)/\omega(\bxi_i)$.  Combining Eqs.~\eqref{eq:f2}
and \eqref{eq:f3}, the \textit{discrete body force} up to the third moment is
\begin{eqnarray}
	\label{eq:fi}
	F_i &=& w_i\rho(\bg\cdot\bxi_i) \left\{1 + \bu\cdot\bxi_i+ \frac 12
	\left[(\bu\cdot\bxi_i)^2 -u^2 + (\theta-1)(\xi^2_i-D-2)\right]\right\}
	-w_i\rho(\bg\cdot\bu)(1+\bu\cdot\bxi_i) \nonumber \\
	&&-\frac{w_i}2\left[(\bg\cdot\bxi_i)\bsigma:\bxi_i\bxi_i
	- 2\bsigma:\bg\bxi_i\right].
\end{eqnarray}
Further discretizing space and time by integrating Eq.~\eqref{eq:dbe} using
second-order schemes~\citep{He1998,Li2022}, we arrive at the lattice Boltzmann
equation with body force
\begin{equation}
	\label{eq:lbm}
	f_i(\bx+\bxi, t+1) - f_i(\bx, t)
	= -\frac{1}{\hat{\tau}} \left[f_i - \overline{\f{0}_i}\right] +
	 	\left[1-\frac{1}{2\hat{\tau}}\right]\overline{F_i},
\end{equation}
where $\hat{\tau} = \tau + 1/2$, $\f{0}_i\equiv w_i\f{0}_N(\bx, \bxi_i,
t)/\omega(\bxi_i)$ with $\f{0}_N$ being the finite Hermite expansion of $\f{0}$.
The over-line stands for evaluation using values of density, velocity and
temperature at the mid-point of a time step.  Noticing the conservation of mass
and heat by both the normal collision operator and the body-force term, these
are $\rho$, $\bu + \bg/2$ and $\theta$ respectively. The deviatoric stress
tensor at the mid-point of a time step, $\overline{\bsigma}$, can be calculated
by
\begin{equation}
    \overline{\bsigma} = -\left(1-\frac{1}{2\hat{\tau}}\right)\left\{\sum_{i}\left[f_i - \overline{\f{0}_i}\right]
    \left(\bxi_i-\bu\right)^2 - \frac{\rho \bg^2}{2}\right\}.
\end{equation}
Nevertheless, in the quasi-equilibrium regime where $f - \f{0} \ll \f{0}$, the
above contribution can be neglected all together as demonstrated by numerical
verifications and Chapman-Enskog analysis in the later sections. We note that
expressions similar to Eq.~\eqref{eq:fi} can be obtained for any order of
expansion to taking into account higher moments of the distribution function.
Particularly, if only the second-moment contribution is retained,
Eq.~\eqref{eq:lbm} is identical to the forcing scheme of~\cite{Guo2002}.

We now analyse the macroscopic effects of the force term by Chapman-Enskog
analysis. As shown previously by~\cite{Shan2006b}, with the Hermite-expanded
distribution function and BGK collision operator, the Chapman-Enskog asymptotic
analysis results in recurrent relations among the Hermite coefficients of
various orders. Particularly for the first approximations (Navier-Stokes), we
have
\begin{equation}
\label{eq:a1}
	\ba{n}_1 = -\tau\left[\frac{\p \ba{n}_0 }{\p t} + \nabla\cdot\ba{n+1}_0
	+ n\nabla\ba{n-1}_0 - n\bg\ba{n-1}_0\right].
\end{equation}
As it is well-known that the equation above leads to the correct
Navier-Stokes-Fourier equations~\citep{Huang1987,Shan2006b}, the error in the
macroscopic equations caused by insufficient expansion of the force term can be
analyzed by examining the contributions of the last term on the \textit{r.h.s.}.
In most force schemes it is retained up to $n\leq 2$. While accurate for the
momentum equation, this causes an error in the third Hermite coefficient of the
first correction $\ba{3}_1 = -3\tau\bg\ba{2}_0$ which in turn results in the
following error in the heat flux
\begin{align}
	\bm{q}_{err} & = \frac 12\int f_{err}c^2\bc d\bxi
	=\frac{1}{12}\ba{3}_1:\int\omega(\bxi)\bH{3}(\bxi)c^2\bc d\bxi\nonumber \\
	             & = -\frac{1}{2}\rho\tau
	\left\{[u^2+(\theta-1)(D+2)]\bg+2(\bg\cdot\bu)\bu\right\},
\end{align}
which is second-order in $u$, first-order in $\theta-1$, and can not be ignored
in flows with strong thermal effects.  However, the non-equilibrium part in
Eq.~\eqref{eq:fi}, $-3\bg\bsigma$, contributes to the heat flux as
\begin{equation}
    \bm{q}_{err} = \frac{1}{2}\tau\bg\cdot\bsigma
    \approx \frac{1}{2}\tau^2p\bg\cdot
    \left[\nabla\bu + \left(\nabla\bu\right)^T -
    	\frac{2}{D}\left(\nabla\cdot\bu\right)\bdelta\right],
\end{equation}
which is $O(\tau^2)$ and can be omitted in continuum flows but may have
significant effect in non-equilibrium ones.

\section{Numerical Validation}

\label{sec:numerical}

To verify the above analysis, we conducted two numerical tests, \textit{i.e.},
compressible Poiseuille flow under cross gravity and heat transfer between two
concentric cylinders under centrifugal force.  These flows are chosen as the
deviatoric stress does not automatically vanish by configuration.  In all LBM
simulations the equilibrium distribution is expanded to fourth order and the
two-dimensional, ninth-order accurate D2Q37 quadrature rule is
employed~\citep{Shan2016}.  Since this model involves discrete velocities of
multiple layers, we have developed a multi-speed mass-conserving boundary
condition (BC) which will be published elsewhere. For the present purposes the
results are insensitive to the implementation of the BC.

\subsection{Compressible Poiseuille flow under cross acceleration}

Consider a two-dimensional steady laminar flow between two infinite horizontal
plates under a homogeneous constant gravity force $\bg=(g_x,g_y)$, where $x$ and
$y$ are respectively the horizontal and vertical directions. The flow is assumed
to be steady and one-dimensional with velocity in the $x$ direction only and all
quantities are functions of $y$. We have therefore $u_y = 0$ and
$\partial/\partial x = \partial/\partial t = 0$. The Navier-Stokes-Fourier
equations reduce to
\begin{equation}
	\label{eq:pois_macro_eq}
	-\mu\frac{\partial^2 u_x}{\partial y^2} = \rho g_x,\quad
	\frac{\p \rho \theta}{\p y} = \rho g_y,\quad\mbox{and}\quad
	\mu\left(\frac{\p u_x}{\p y}\right)^2 + \lambda \frac{\p^2 \theta}{\p y^2} = 0,
\end{equation}
where $\rho$, $u_x$ and $\theta$ are density, velocity and temperature
respectively, $\mu$ and $\lambda$ the dynamic viscosity and heat conductivity,
both assumed constant. Dirichlet boundary conditions are imposed for $u_x$ and
$\theta$ on the bottom and top boundaries at $y=0$ and $H$ respectively
\begin{equation}
	u_x|_{y=0} = U_b, \quad u_x|_{y=H} = U_t,\quad
	\theta|_{y=0} = \theta_b, \quad \theta|_{y=H} = \theta_t,
\end{equation}
where $U_b$, $U_t$ and $\theta_b$, $\theta_t$ are the tangential velocity and
temperature at the bottom and top boundaries respectively. As
Eqs.~\eqref{eq:pois_macro_eq} are non-trivial to solve analytically, for the
reference solution, we utilized a finite-difference scheme which is detailed in
Appendix~\ref{sec:fd}.

All LBM simulations of compressible Poiseuille flow were performed on a
$L_x\times L_y$ lattice where $L_x = 3$ and $L_y = 150$. The channel hight is
thus $H=L_yc$ where $c \approx 1.19698$ is the lattice constant of
D2Q37~\citep{Shan2016}. The velocity at both walls were set at $U_t=U_b=0$. To
investigate the effect of temperature gradient, two simulations were performed
for $\theta_b = 1.0$, $\theta_t = 1.1$ and $\theta_b = 0.7$, $\theta_t = 1.4$,
corresponding to a total cross-channel temperature variation of $10\%$ and
$100\%$ respectively.  Define $U_c = \rho_0g_xH^2/8\mu$ which is the velocity
(and also the Mach number \textit{w.r.t.}\ the isothermal speed of sound) at the
centre of channel when both $\rho$ and $\theta$ are homogeneous. The
corresponding Reynolds number is $Re = \rho_0U_cH/\mu$ where $\rho_0$ is the
averaged density.  Setting $U_c = 1.5$, $Re = 1800$ and $\rho_0 = 1$, the other
parameters are determined as: $g_x = U^2_c/Re H$, $\mu = \rho_0U_cH/Re$ and
$\lambda = c_p\mu/ Pr$, where the Prandtl number $Pr=1$ for BGK collision term,
and the isobaric heat capacity $c_p = (D+2)/2$. To maintain both $\mu$ and
$\lambda$ as homogeneous constants, the relaxation time was set to $\tau =
\mu/\rho\theta$.  The cross-flow gravity, $g_y$, points downward and was set so
that $g_y/g_x = -50$.

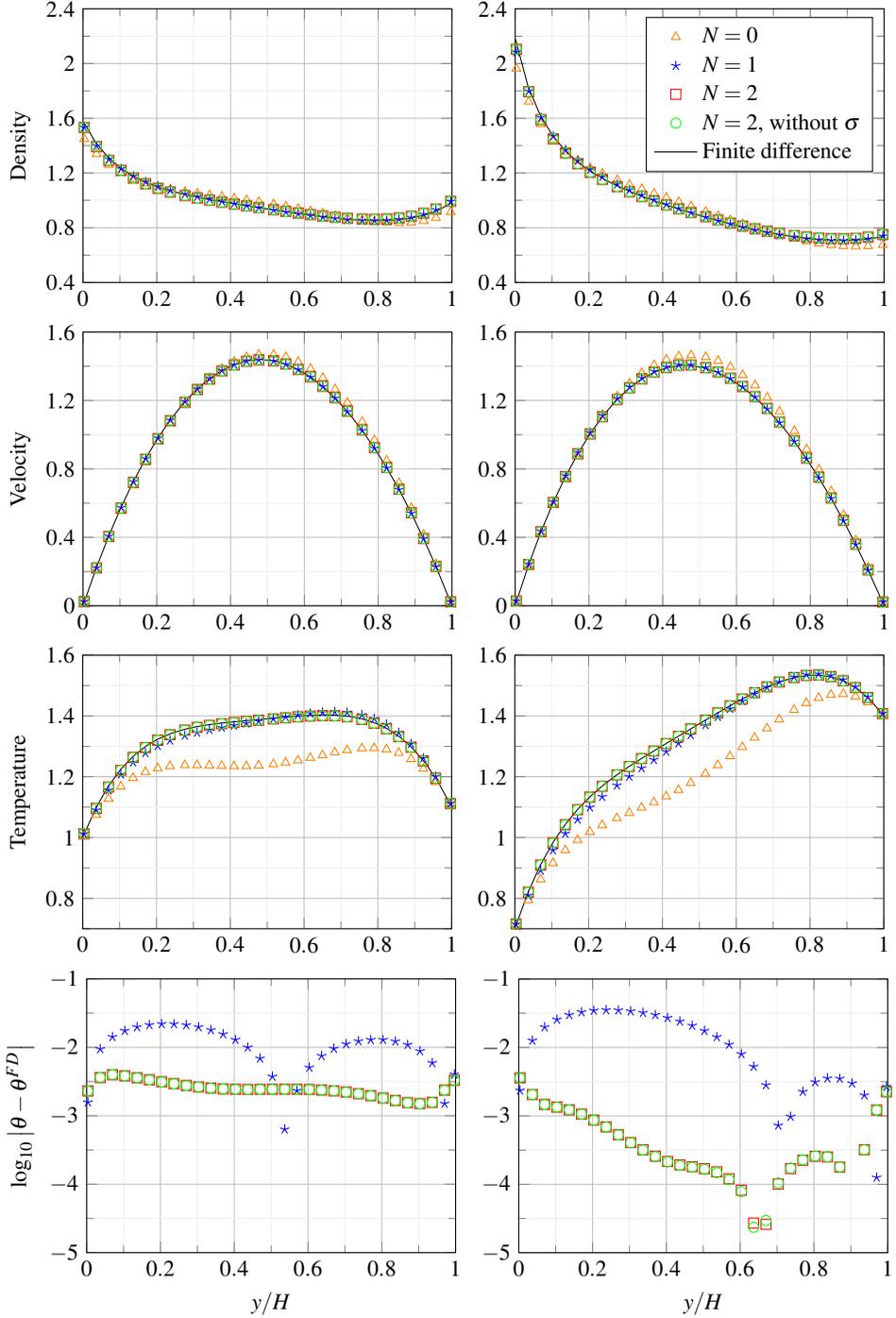
\begin{figure}
\begin{center}
\begin{tikzpicture}
	\begin{axis}[
		xmin = 0, xmax = 1,
		ymin = 0.4, ymax = 2.4,
		xtick distance = 0.2,
		ytick distance = 0.4,
		grid = both,
		minor tick num = 1,
		major grid style = {lightgray},
		minor grid style = {lightgray!25},
		width = 0.5\textwidth,
		height = 0.4*\textwidth,
		legend cell align = {left},
		legend pos=north east,
		ylabel = {Density}
		]
		\addplot[orange, mark= triangle, only marks] table {pois_dtheta=0.1_f1_rho.dat};
		\addplot[blue, mark= star, only marks] table {pois_dtheta=0.1_f2_rho.dat};
		\addplot[red, mark= square, only marks] table {pois_dtheta=0.1_f3_rho.dat};
		\addplot[green, mark= o, only marks] table {pois_dtheta=0.1_f30_rho.dat};
		\addplot[black] table {pois_dtheta=0.1_fd_rho.dat};
	\end{axis}	
\end{tikzpicture}
\begin{tikzpicture}
	\begin{axis}[
		xmin = 0, xmax = 1,
		ymin = 0.4, ymax = 2.4,
		xtick distance = 0.2,
		ytick distance = 0.4,
		grid = both,
		minor tick num = 1,
		major grid style = {lightgray},
		minor grid style = {lightgray!25},
		width = 0.5\textwidth,
		height = 0.4*\textwidth,
		legend cell align = {left},
		legend pos=north east
		]
		\addplot[orange, mark= triangle, only marks] table {pois_dtheta=0.7_f1_rho.dat};
		\addplot[blue, mark= star, only marks] table {pois_dtheta=0.7_f2_rho.dat};
		\addplot[red, mark= square, only marks] table {pois_dtheta=0.7_f3_rho.dat};
		\addplot[green, mark= o, only marks] table {pois_dtheta=0.7_f30_rho.dat};
		\addplot[black] table {pois_dtheta=0.7_fd_rho.dat};
		\legend{$N=0$, $N=1$, $N=2$, {$N=2$, without $\sigma$}, Finite difference}
	\end{axis}
\end{tikzpicture}
\begin{tikzpicture}
	\begin{axis}[
		xmin = 0, xmax = 1,
		ymin = 0, ymax = 1.6,
		xtick distance = 0.2,
		ytick distance = 0.4,
		grid = both,
		minor tick num = 1,
		major grid style = {lightgray},
		minor grid style = {lightgray!25},
		width = 0.5\textwidth,
		height = 0.4*\textwidth,
		legend cell align = {left},
		legend style={at={(0.65,0.6)}},
		ylabel = {Velocity}
		]
		\addplot[orange, mark= triangle, only marks] table {pois_dtheta=0.1_f1_u.dat};
		\addplot[blue, mark= star, only marks] table {pois_dtheta=0.1_f2_u.dat};
		\addplot[red, mark= square, only marks] table {pois_dtheta=0.1_f3_u.dat};
		\addplot[green, mark= o, only marks] table {pois_dtheta=0.1_f30_u.dat};
		\addplot[black] table 
		{pois_dtheta=0.1_fd_u.dat};
	\end{axis}
\end{tikzpicture}
\begin{tikzpicture}
	\begin{axis}[
		xmin = 0, xmax = 1,
		ymin = 0, ymax = 1.6,
		xtick distance = 0.2,
		ytick distance = 0.4,
		grid = both,
		minor tick num = 1,
		major grid style = {lightgray},
		minor grid style = {lightgray!25},
		width = 0.5\textwidth,
		height = 0.4*\textwidth,
		legend cell align = {left},
		legend style={at={(0.65,0.6)}}
		]
		\addplot[orange, mark= triangle, only marks] table {pois_dtheta=0.7_f1_u.dat};
		\addplot[blue, mark= star, only marks] table {pois_dtheta=0.7_f2_u.dat};
		\addplot[red, mark= square, only marks] table {pois_dtheta=0.7_f3_u.dat};
		\addplot[green, mark= o, only marks] table {pois_dtheta=0.7_f30_u.dat};
		\addplot[black] table 
		{pois_dtheta=0.7_fd_u.dat};
	\end{axis}
\end{tikzpicture}
\begin{tikzpicture}
\begin{axis}[
    xmin = 0, xmax = 1,
    ymin = .7, ymax = 1.6,
    xtick distance = 0.2,
    ytick distance = 0.2,
    grid = both,
    minor tick num = 1,
    major grid style = {lightgray},
    minor grid style = {lightgray!25},
    width = 0.5\textwidth,
    height = 0.4*\textwidth,
    legend cell align = {left},
    legend style={at={(0.7,0.7)}},
    ylabel = {Temperature}
]
    \addplot[orange, mark= triangle, only marks] table {pois_dtheta=0.1_f1_theta.dat};
    \addplot[blue, 	 mark= star, 	 only marks] table {pois_dtheta=0.1_f2_theta.dat};
    \addplot[red, 	 mark= square, 	 only marks] table {pois_dtheta=0.1_f3_theta.dat};
    \addplot[green,  mark= o, 		 only marks] table {pois_dtheta=0.1_f30_theta.dat};
    \addplot[black] table 
    {pois_dtheta=0.1_fd_theta.dat};
\end{axis}
\end{tikzpicture}
\begin{tikzpicture}
\begin{axis}[
	xmin = 0, xmax = 1,
	ymin = 0.7, ymax = 1.6,
	xtick distance = 0.2,
	ytick distance = 0.2,
	grid = both,
	minor tick num = 1,
	major grid style = {lightgray},
	minor grid style = {lightgray!25},
	width = 0.5\textwidth,
    height = 0.4*\textwidth,
	legend cell align = {left},
	legend style={at={(0.65,0.6)}}
	]
	\addplot[orange, mark=triangle, only marks] table {pois_dtheta=0.7_f1_theta.dat};
	\addplot[blue, 	 mark=star, 	only marks] table {pois_dtheta=0.7_f2_theta.dat};
	\addplot[red, 	 mark=square, 	only marks] table {pois_dtheta=0.7_f3_theta.dat};
	\addplot[green,  mark=o, 		only marks] table {pois_dtheta=0.7_f30_theta.dat};
	\addplot[black] table {pois_dtheta=0.7_fd_theta.dat};
\end{axis}
\end{tikzpicture}
\begin{tikzpicture}
	\begin{axis}[
		xmin = 0, xmax = 1,
		ymin = -5, ymax = -1,
		grid = both,
		minor tick num = 1,
		major grid style = {lightgray},
		minor grid style = {lightgray!25},
		width = 0.5\textwidth,
	    height = 0.4*\textwidth,
		legend cell align = {left},
		legend pos=north east,
		ylabel = {$\log_{10}\left|\theta-\theta^{FD}\right|$},
		xlabel = {$y/H$}
		]
		\addplot[blue, mark= star, only marks] table {pois_dtheta=0.1_f2_theta_e.dat};
		\addplot[red, mark= square, only marks] table {pois_dtheta=0.1_f3_theta_e.dat};
		\addplot[green, mark= o, only marks] table {pois_dtheta=0.1_f30_theta_e.dat};
		
	\end{axis}	
\end{tikzpicture}
\begin{tikzpicture}
	\begin{axis}[
		xmin = 0, xmax = 1,
		ymin = -5, ymax = -1,
		grid = both,
		minor tick num = 1,
		major grid style = {lightgray},
		minor grid style = {lightgray!25},
		width = 0.5\textwidth,
	    height = 0.4*\textwidth,
		legend cell align = {left},
		legend pos=north east,
		xlabel = {$y/H$}
		]
		\addplot[blue, mark= star, only marks] table {pois_dtheta=0.7_f2_theta_e.dat};
		\addplot[red, mark= square, only marks] table {pois_dtheta=0.7_f3_theta_e.dat};
		\addplot[green, mark= o, only marks] table {pois_dtheta=0.7_f30_theta_e.dat};
	\end{axis}	
\end{tikzpicture}
\end{center}

\caption{Compressible Poiseuille flow with cross-flow gravity and heat gradient.
	Shown from top to bottom are profiles of density, velocity, temperature, and
	errors in temperature as computed by the force term expanded to orders
	corresponding to $N=0$, $1$ and $2$ in Eq.~\eqref{eq:F}. The reference is a
	high-precision finite-difference solution which is also shown. In the left
	column are the results for $\theta_b = 1.0$, $\theta_t = 1.1$, corresponding to
	a 10\% total temperature variation, and in the right column are those for
	$\theta_b = 0.7$, $\theta_t = 1.4$, corresponding to a 100\% total temperature
	variation.}

\label{fig:figure 1}
\end{figure}

Shown in Figs.~\ref{fig:figure 1} are the profiles of density, velocity,
temperature, and errors in temperature for the two sets of simulations with
different temperature variations. To demonstrate the effects of the various
moments in the force term, simulations were performed with the force term
expanded to three different orders, corresponding to $N = 0$, $1$, and $2$ in
Eq.~\eqref{eq:F}.  In addition, the cases of $N=2$ were run with and without the
contributions of $\bsigma$ in Eq.~\eqref{eq:fi}.  As the reference solution, a
high-resolution ($N=450$) finite-difference results are also shown. We first
note that due to the presence of cross-flow gravity and temperature gradient,
the density and temperature are asymmetric \textit{w.r.t.}\ to the centre line.
The velocity peak is also less than $U_c$ and does not occur at the centre.
These effects are stronger in the cases with larger temperature variation.

Consistent with the theoretical analysis that the second-order ($N=1$) force
term is necessary in recovering the momentum equation and conservation of
internal energy, it is evident from simulation results that retaining only the
first-order force term ($N=0$) causes significant overall discrepancies. The
more common second-order ($N=1$) approximation yields satisfactory results for
density and velocity, in accordance with the theoretical prediction of
Eq.~\eqref{eq:a1}. However, noticeable differences between the second-order and
reference solutions appear in the temperature profile and are more pronounced in
the high temperature-variation cases. This discrepancy is eliminated in the
third-order ($N=2$) solutions, confirming the correctness and necessity of the
third-order contribution by Eq.~\eqref{eq:fi}.  Moreover, the contribution of
the deviatoric tensor to the force term appears to be negligible in all cases
tested, confirming the Chapman-Enskog analysis in the previous section.

\subsection{Heat transfer between concentric cylinders under centrifugal force}

The second test is the two-dimensional heat transfer between two concentric
cylinders maintained at different temperatures and rotating with the same
angular speed, $\alpha$.  In the rotating reference frame, the fluid is assumed
static and subject to the centripetal acceleration of $\alpha^2R$ where $R$ is
the radial coordinate.  The Navier-Stokes-Fourier equations have the following
one-dimensional solution
\begin{equation}
 	\theta = \theta_i + (\theta_{o}-\theta_i)\frac{\ln{R/R_i}}{\ln{R_o/R_i}},
 	\quad\mbox{and}\quad
	\rho = \frac{\rho_i\theta_i}{\theta}\exp\left[\int_{R_i}^R\frac{\alpha^2r}{\theta(r)}dr\right],
\end{equation}
where $\rho$, $\theta$ are density and temperature respectively, the subscripts
$i$ and $o$ denote values at the inner and outer cylinders. The reference
solution of  density is solved by a high-precision numerical integration.
Required the average of density $\bar{\rho} = \rho_0$, the $\rho_i$ is obtained
by integrating the density in the whole domain.

The LBM simulations were performed on a $L\times L$ lattice where $L=250$. The
radial centripetal acceleration is $-\alpha^2R$. The radius ratio $R_o/R_i$ is
set to 5 with $R_o=Lc/2$. Using the width of the annular, $R_o-R_i$, as the
characteristic length and the centre-line velocity, $U_c\equiv\alpha(R_o +
R_i)/2$, as the characteristic speed, the Reynolds number is defined as $Re =
\rho_0\alpha(R_o^2-R_i^2)/2\mu$ where $\mu$ is the dynamic viscosity.  As in the
previous case, the relaxation time is set to $\tau=\mu/\rho\theta$ so that $\mu$
is a constant.  Note that the fluid is static, and the centre-line velocity,
$U_c$, measures the compressibility as the isothermal Mach number, and the
Reynolds number only plays the role of a dimensionless viscosity. As there is no
flow in the rotating frame, all quantities are insensitive to the Reynolds
number.

\begin{figure}
\begin{center}
\begin{tikzpicture}
	\begin{axis}[
		xmin = 0, xmax = 1,
		grid = both,
		minor tick num = 1,
		major grid style = {lightgray},
		minor grid style = {lightgray!25},
		width = 0.47\textwidth,
		legend cell align = {left},
		legend style={at={(0.75,1.)}} ,
		ylabel = {$\rho$}
		]
		\addplot[orange, mark= triangle, only marks] table {heat_f1_rho_v.dat};
		\addplot[blue, mark= star, only marks] table {heat_f2_rho_v.dat};
		\addplot[red, mark= square, only marks] table {heat_f3_rho_v.dat};
		\addplot[green, mark= o, only marks] table {heat_f30_rho_v.dat};
		\addplot[black] table {heat_a_rho_v.dat};
		\legend{$N=0$, $N=1$, $N=2$, {$N=2$, without $\sigma$}, Analytical}
	\end{axis}
\end{tikzpicture} 
\begin{tikzpicture}
	\begin{axis}[
		xmin = 0, xmax = 1,
		grid = both,
		minor tick num = 1,
		major grid style = {lightgray},
		minor grid style = {lightgray!25},
		width = 0.47\textwidth,
		legend cell align = {left},
		legend style={at={(0.75,1.)}} ,
		]
		\addplot[orange, mark= triangle, only marks] table {heat_dtheta=0.7_f1_rho.dat};
		\addplot[blue, mark= star, only marks] table {heat_dtheta=0.7_f2_rho.dat};
		\addplot[red, mark= square, only marks] table {heat_dtheta=0.7_f3_rho.dat};
		\addplot[green, mark= o, only marks] table {heat_dtheta=0.7_f30_rho.dat};
		\addplot[black] table {heat_dtheta=0.7_a_rho.dat};
	\end{axis}
\end{tikzpicture} 

\begin{tikzpicture}
	\begin{axis}[
		xmin = 0, xmax = 1,
		ymin = -6, ymax = -1,
		ytick distance = 1,
		grid = both,
		minor tick num = 1,
		major grid style = {lightgray},
		minor grid style = {lightgray!25},
		width = 0.47\textwidth,
		legend cell align = {left},
		legend pos=north east,
		ylabel = {$\log_{10}\left|\rho-\rho^{a}\right|$}
		]
		\addplot[orange, mark= triangle, only marks] table {heat_f1_rho_e.dat};
		\addplot[blue, mark= star, only marks] table {heat_f2_rho_e.dat};
		\addplot[red, mark= square, only marks] table {heat_f3_rho_e.dat};
		\addplot[green, mark= o, only marks] table {heat_f30_rho_e.dat};
	\end{axis}
\end{tikzpicture}
\begin{tikzpicture}
	\begin{axis}[
		xmin = 0, xmax = 1,
		ymin = -6, ymax = -1,
        ytick distance = 1,
		grid = both,
		minor tick num = 1,
		major grid style = {lightgray},
		minor grid style = {lightgray!25},
		width = 0.47\textwidth,
		legend cell align = {left},
		legend style={at={(0.75,1.)}} 
		]
		\addplot[orange, mark= triangle, only marks] table {heat_dtheta=0.7_f1_rho_e.dat};
		\addplot[blue, mark= star, only marks] table {heat_dtheta=0.7_f2_rho_e.dat};
		\addplot[red, mark= square, only marks] table {heat_dtheta=0.7_f3_rho_e.dat};
		\addplot[green, mark= o, only marks] table {heat_dtheta=0.7_f30_rho_e.dat};
	\end{axis}
\end{tikzpicture} 

\begin{tikzpicture}
	\begin{axis}[
		xmin = 0, xmax = 1,
		grid = both,
		minor tick num = 1,
		major grid style = {lightgray},
		minor grid style = {lightgray!25},
		width = 0.47\textwidth,
		legend cell align = {left},
		legend pos=north east,
		ylabel = {$\theta$}
		]
		\addplot[orange, mark= triangle, only marks] table {heat_f1_theta_v.dat};
		\addplot[blue, mark= star, only marks] table {heat_f2_theta_v.dat};
		\addplot[red, mark= square, only marks] table {heat_f3_theta_v.dat};
		\addplot[green, mark= o, only marks] table {heat_f30_theta_v.dat};
		\addplot[black] table {heat_a_theta_v.dat};
	\end{axis}
\end{tikzpicture}
\begin{tikzpicture}
	\begin{axis}[
		xmin = 0, xmax = 1,
		grid = both,
		minor tick num = 1,
		major grid style = {lightgray},
		minor grid style = {lightgray!25},
		width = 0.47\textwidth,
		legend cell align = {left},
		legend pos=north east,
		]
		\addplot[orange, mark= triangle, only marks] table {heat_dtheta=0.7_f1_theta.dat};
		\addplot[blue, mark= star, only marks] table {heat_dtheta=0.7_f2_theta.dat};
		\addplot[red, mark= square, only marks] table {heat_dtheta=0.7_f3_theta.dat};
		\addplot[green, mark= o, only marks] table {heat_dtheta=0.7_f30_theta.dat};
		\addplot[black] table {heat_dtheta=0.7_a_theta.dat};
	\end{axis}
\end{tikzpicture}

   \begin{tikzpicture}
	\begin{axis}[
		xmin = 0, xmax = 1,
		ymin = -6, ymax = -1,
        ytick distance = 1,
		grid = both,
		minor tick num = 1,
		major grid style = {lightgray},
		minor grid style = {lightgray!25},
		width = 0.47\textwidth,
		legend cell align = {left},
		legend pos=north east,
		xlabel={$\left(R-R_i\right)/(R_o-R_i)$},
		ylabel = {$\log_{10}\left|\theta-\theta^a\right|$}
		]
		\addplot[orange, mark= triangle, only marks] table {heat_f1_theta_e.dat};
		\addplot[blue, mark= star, only marks] table {heat_f2_theta_e.dat};
		\addplot[red, mark= square, only marks] table {heat_f3_theta_e.dat};
		\addplot[green, mark= o, only marks] table {heat_f30_theta_e.dat};
	\end{axis}
\end{tikzpicture} 
\begin{tikzpicture}
	\begin{axis}[
		xmin = 0, xmax = 1,
		ymin = -6, ymax = -1,
        ytick distance = 1,
		grid = both,
		minor tick num = 1,
		major grid style = {lightgray},
		minor grid style = {lightgray!25},
		width = 0.47\textwidth,
		legend cell align = {left},
		legend pos=north east,
		xlabel={$\left(R-R_i\right)/(R_o-R_i)$},
		]
		\addplot[orange, mark= triangle, only marks] table {heat_dtheta=0.7_f1_theta_e.dat};
		\addplot[blue, mark= star, only marks] table {heat_dtheta=0.7_f2_theta_e.dat};
		\addplot[red, mark= square, only marks] table {heat_dtheta=0.7_f3_theta_e.dat};
		\addplot[green, mark= o, only marks] table {heat_dtheta=0.7_f30_theta_e.dat};
	\end{axis}
\end{tikzpicture}

\end{center}

\caption{Heat conduction between two rotating concentric cylinders. Shown are
	the radial profiles of density and temperature and the associated errors
	computed with the force terms expanded to various orders. The superscript $a$
	denotes the analytical solutions. The left column shows the result for
	$\theta_i=1$, $\theta_o=1.1$, and the right column shows the result for
	$\theta_i=1.4$, $\theta_o=0.7$.}

\label{fig:figure 2}
\end{figure}
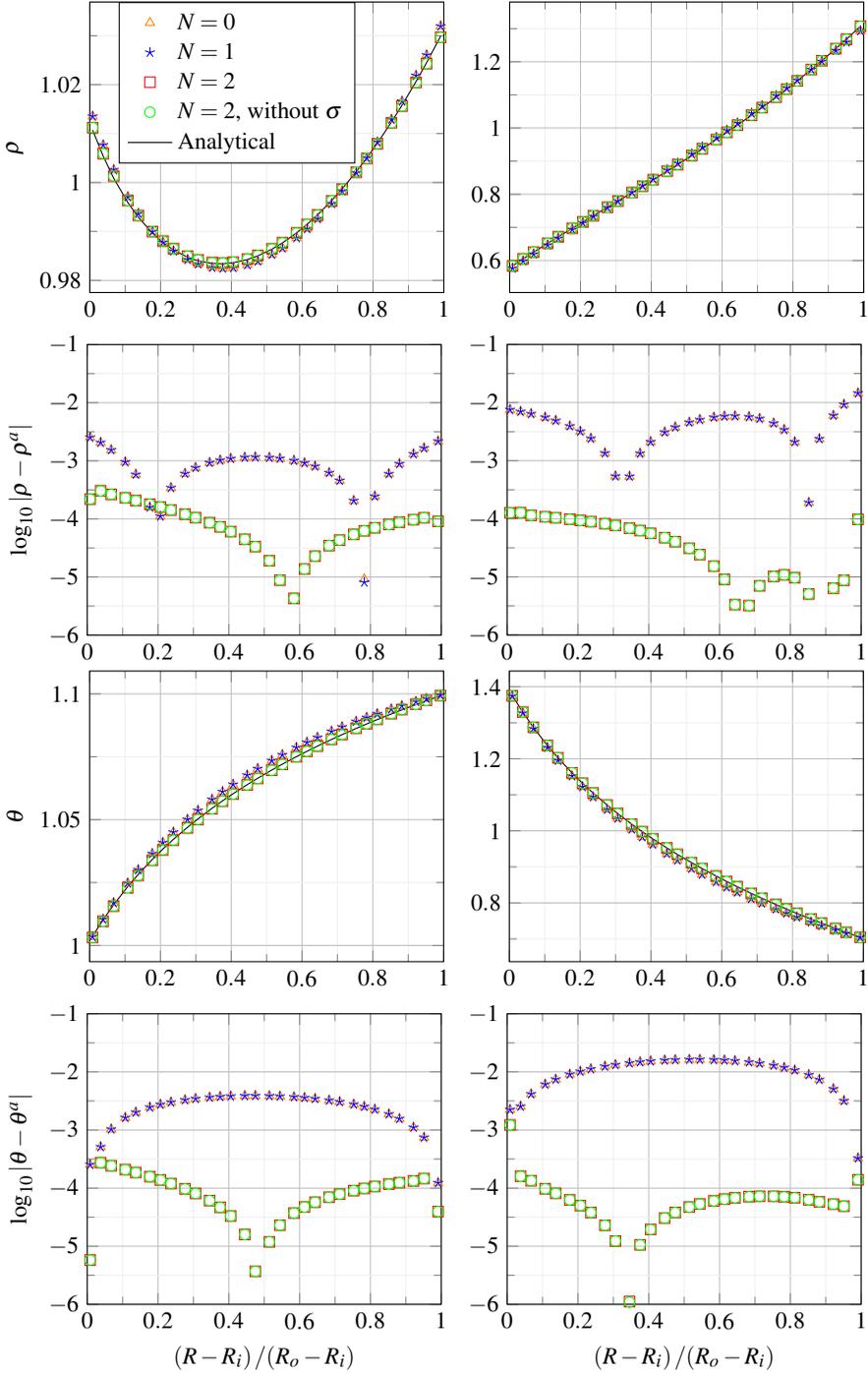

Shown in Figs.~\ref{fig:figure 2} are the radial profiles of density,
temperature and the associated errors as computed with the force term expanded
to various orders. The parameters are: $Re = 100$, $U_c = 0.3$, and $\rho_0=1$.
Two simulations were performed for $\theta_i=1$, $\theta_o=1.1$, and for
$\theta_i=1.4$, $\theta_o=0.7$.  Note that the first and second expansions are
identical as $\bu = 0$ in this case. To be seen is that if the third-moment
contribution is omitted, the errors in temperature profile is significant
especially in the larger temperature variation case.  It is also evident from
the results that the third-order contribution is necessary to correctly recover
the energy equation. The contribution of the stress tensor, $\bsigma$, vanishes
as $\bsigma$ itself vanishes in absence of fluid flow.



\section{Conclusion}

\label{sec:discussion}

In summary, we gave a generic approach of incorporating body force in lattice
Boltzmann method based on the Hermite expansion of the force term in the
Boltzmann equation. In particular, a novel LBM forcing scheme for thermal
compressible flows with second-order time accuracy is obtained which includes
the third-order contribution of the force term.  The errors caused by the
omission of this correction in common practices is identified with an error in
heat flux \textit{via} Chapman-Enskog analysis. All theoretical findings,
including the correctness and necessity of the new forcing term, are confirmed
numerically.

We note that the present approach is independent of neither the form of the
equilibrium distribution nor the underlying lattice, and can be
straightforwardly extended to include higher moments which are significant in
non-equilibrium flows.  Although for simplicity, we used the BGK collision model
in present work.  It is also straightforward to adapt it to more complicated
collision models such as the spectral multiple relaxation time model.

\appendix

\section{Finite-difference solution of compressible Poiseuille flow}

\label{sec:fd}

We outline the one-dimensional finite-difference solution of the steady,
gravity-driven, compressible Poiseuille flow with cross-flow gravity and
temperature gradient. Along the channel height the domain is divided into $N$
nodes with $\Delta y = H/(N-1)$. Let $\rho_j, u_j,\theta_j$ be the density,
velocity and temperature at the $j$-th node. Applying second-order central
difference scheme to the horizontal momentum and energy equations, and first
order forward scheme to the vertical momentum equation,
Eqs.~\eqref{eq:pois_macro_eq} are discretized as:
\begin{subequations}
\begin{align}
    u_{j+1}-2u_j+u_{j-1} &=-\frac{g_x\Delta y^2}{\mu} \rho_j \quad &2\leq j\leq N-1 ,\\
    \rho_{j+1}\theta_{j+1} - \rho_j\theta_j  &= \rho_jg_y\Delta y  \quad   &1\leq j\leq N-1,\\
    \mu\left(u_{j+1}-u_{j-1}\right)^2 + 4\lambda\left(\theta_{j+1}-2\theta_{j}+\theta_{j-1}\right) &=0  \quad &2\leq j\leq N-1.
\end{align}
\end{subequations}
The Dirichlet boundary condition dictates that
\begin{equation}
  	u_1 = U_b, \quad u_N = U_t,\quad
	\theta_1 = \theta_b, \quad \theta_N = \theta_t.
\end{equation}
The mass conservation yields
\begin{equation}
   \left( \frac{\rho_1}{2} + \sum_{j=2}^{N-1}\rho_j + \frac{\rho_N}{2}\right)
   = \rho_0(N-1).
\end{equation}
There are $3N-2$ equations for $3N-2$ unknowns. As the above discretized
equations are non-linear and coupled, a iterative method is used where the
non-linear term $\mu(u_{j+1}-u_{j-1})^2$ is evaluated using variables at the
previous iteration. The equations at iteration $t$ can be written as:
\begin{subequations}
\label{eq:iter_eq}
\begin{align}
    u_{j+1}^{t}-2u_j^t+u_{j-1}^t + \frac{g_x\Delta y^2}{\mu}\rho^t_j &=0, \quad &2\leq j\leq N-1, \label{eq:d_u_eq}\\
    \rho_{j+1}^t\theta_{j+1}^t - \rho_j^t\left(\theta_j^t + g_y\Delta y \right)&= 0,  \quad   &1\leq j\leq N-1,
    \label{eq:d_p_eq}\\
    \mu\left(u^{t-1}_{j+1}-u^{t-1}_{j-1}\right)^2 + 4\lambda\left(\theta_{j+1}^t-2\theta^t_{j}+\theta_{j-1}^t\right) &=0 , \quad &2\leq j\leq N-1,
    \label{eq:d_energy}\\
    \left( \frac{\rho_1^t}{2}  + \sum_{j=2}^{N-1}\rho_j^t + \frac{\rho_N^t}{2}\right) &= \rho_0(N-1)\label{eq:d_mass}.
\end{align}
\end{subequations}
The Eq.~\eqref{eq:iter_eq} is a system of linear equations which was solved by
the following process. With the initial guess of $u$ chosen as:
\begin{equation}
    u = -\frac{g_x}{2\mu}y(y-H) + \frac{U_t-U_b}{H} y+U_b, \quad y\in[0,H].
\end{equation}
the temperature field can be solved from Eq.~\eqref{eq:d_energy}. The density
field is solved next from Eqs.~\eqref{eq:d_p_eq}--\eqref{eq:d_mass}, followed by
solving the velocity field from Eq.~\eqref{eq:d_u_eq}. This process is iterated
until convergence.

\section*{Acknowledgment}

This work was supported by the National Natural Science Foundation of China
Grant 92152107, Dept.\ of Science and Technology of Guangdong Province Grant
2020B1212030001, and Shenzhen Science and Technology Program Grant
KQTD20180411143441009.
\section*{Declaration of Interests}
The authors report no conflict of interest.

\bibliographystyle{jfm}
\bibliography{force_expansion}

\end{document}